\documentclass{book}
\usepackage{Wires}
\usepackage[dvips,pdfmark]{hyperref}
\usepackage{endfloat}

\begin{document}
\Overview

\arttitle{Identification of significant features in DNA microarray data}

\artid{2DPP}

\fname{Eric}
\sname{Bair}

\affil{Departments of Endodontics and Biostatistics \\
  Univ. of North Carolina at Chapel Hill \\
  Chapel Hill, NC 27599}


\begin{abstract}
DNA microarrays are a relatively new technology that can
simultaneously measure the expression level of thousands of
genes. They have become an important tool for a wide variety of
biological experiments. One of the most common goals of DNA microarray
experiments is to identify genes associated with biological processes
of interest. Conventional statistical tests often produce poor results
when applied to microarray data due to small sample sizes, noisy data,
and correlation among the expression levels of the genes. Thus, novel
statistical methods are needed to identify significant genes in DNA
microarray experiments. This article discusses the challenges inherent
in DNA microarray analysis and describes a series of statistical
techniques that can be used to overcome these challenges. The problem
of multiple hypothesis testing and its relation to microarray studies
is also considered, along with several possible solutions.
\end{abstract}

\begin{introtext}
High-dimensional biological data sets have become increasingly common
in recent years. Examples include data collected from DNA microarrays,
comparative genome hybridization experiments, mass spectrometry,
genome-wide association studies, and DNA/RNA sequencing. These new
technologies have revolutionized our understanding of the genetics of
human disease and numerous other biological processes. However,
statistical analysis of such data sets is challenging for several
reasons. These data sets are high-dimensional, and the sample sizes
are often small. Moreover, many of these data sets tend to be
``noisy,''  and the correlation between the features that are measured
can be complex. For these reasons conventional statistical methods
often produce unsatisfactory results when applied to modern
high-dimensional biological data.

The present study focuses on one of the most common problems in the
analysis of high-dimensional biological data, which is the
identification of significant genes in DNA microarray studies. This is
one of the best-studied problems in the analysis of high-dimensional
biological data sets, and many of the methods that are applied to this
problem may also be applied to other types of high-dimensional
biological data. In a typical microarray study, one may wish to
identify genes that are associated with a disease or some other
biological process of interest. For example, one might attempt to
identify genes associated with a disease by collecting a set of
biological samples from diseased patients and another set of samples
from healthy patients. Genes whose expression levels differ between
the diseased samples and the control samples may be associated with
the disease of interest. Alternatively, one might wish to identify
genes that may be used to predict the prognosis of patients with a
specific type of cancer. One might identify such genes by collecting
tumor samples from a cohort of cancer patients and searching for genes
whose expression levels are associated with the survival times of the
patients. Ultimately, this information may be used for personalized
treatment of cancer and other diseases. If the gene expression profile
of a tumor indicates that the risk of metastasis is high, then the
cancer should be treated more aggressively than another tumor whose
gene expression suggests a low risk of metastasis.

This article consists of three main sections. In the first section, we
will briefly describe DNA microarray technology and how DNA microarray
data is collected. In the second section, we will provide a brief
overview of some of the methods that have been used to identify
significant genes in DNA microarray experiments. Numerous methods have
been proposed in recent years, and space does not permit a detailed
discussion of all possible methods. We have attempted to focus on
several of the most commonly used approaches, along with an overview
of some of the common principles and techniques used in these
methods. We also briefly describe a few more recent methods for
combining information across genes. In the final section, we discuss
the problem of multiple hypothesis testing, which inevitably arises
when identifying significant features in high-dimensional data sets.
\end{introtext}

\section{DNA Microarray Data}
\subsection{Overview of Molecular Biology}
Each organism's genetic information is contained in a molecule called
deoxyribonucleic acid, more commonly known as DNA. DNA is a
double-stranded molecule that is a chain of four possible nucleotides,
namely adenine (A), cytosine (C), guanine (G), and thymine (T). The
two strands of DNA are joined to one another by hydrogen bonds between
nucleotides on the opposite strands. A always pairs with T, and G
always pairs with C. Thus, if the sequence of one strand of DNA is
known, then the sequence of the other stand is also known. Each such
pair of bonded nucleotides is known as a base pair.

There are approximately 3.2 billion base pairs in the human genome
(i.e. the entire sequence of DNA in a given human cell)
\citep{HG04}. Different segments of DNA perform different functions,
and much of the DNA performs no known function. The DNA segments of
primary interest in most studies are the segments which contain
instructions for building proteins. These segments are known as genes,
and they comprise about 1.5\% of the DNA sequence in humans
\citep{eL01}. Proteins perform most of the important functions in
cells, including metabolism, DNA replication and repair, and
communication with other cells.

The information contained in DNA is converted to proteins in a
two-step process: In the first step, known as transcription, a given
sequence of DNA is transcribed into an intermediary called messenger
ribonucleic acid (mRNA), which is a single-stranded molecule that
contains a copy of the complements of the base sequence of the
DNA. The one difference is that thymine is replaced by uracil (U). In
the second step, known as translation, the sequence of base pairs in
the mRNA is translated into a protein, which is composed of a sequence
of amino acids. Each set of three base pairs in the mRNA corresponds
to one of 20 amino acids, a relationship that is known as the genetic
code. This process by which the information in the sequence of DNA is
converted to mRNA and then to proteins is known as the fundamental
dogma of molecular biology. See \citet{sD02} for a discussion of this
process and its relationship to DNA microarray data.

\subsection{DNA Microarray Technology}
An important implication of the fundamental dogma of molecular biology
is that there should be a strong association between the presence of a
given protein in a cell and the presence of the mRNA sequence that is
transcribed to build that protein. If a protein is active in a given
cell, there should be a large number of copies of the mRNA sequence
corresponding to that protein. Conversely, if a protein is not active
in a cell, there should be few copies of the corresponding mRNA
sequence. Thus, DNA microarrays attempt to evaluate the presence or
absence of proteins in a cell and their relative abundance by 
measuring the relative abundance of the corresponding mRNA sequences.

DNA microarrays measure the relative abundance of mRNA sequences in
the cells in a sample by taking advantage of complementary base
pairing. Recall that in a DNA (or RNA) sequence, C always pairs with G
and A always pairs with T (or U). A DNA microarray is typically
constructed by placing an array of probes on a glass microscope
slide. Each probe consists of a sequence of nucleotides that is
complementary to the nucleotide sequence of a specific mRNA or its
corresponding DNA sequence. Thus, one can measure the expression level
of a given gene by measuring the amount of mRNA that hybridizes to the
spot on the microarray corresponding to the gene. Different forms of
DNA microarrays exist, such as oligonucleotide microarrays
\citep{dL96} and cDNA microarrays \citep{jD96, tH01}, but all of the
most commonly used microarrays are based on this principle.

\begin{figure}
  \centerline{\includegraphics[width=\textwidth]{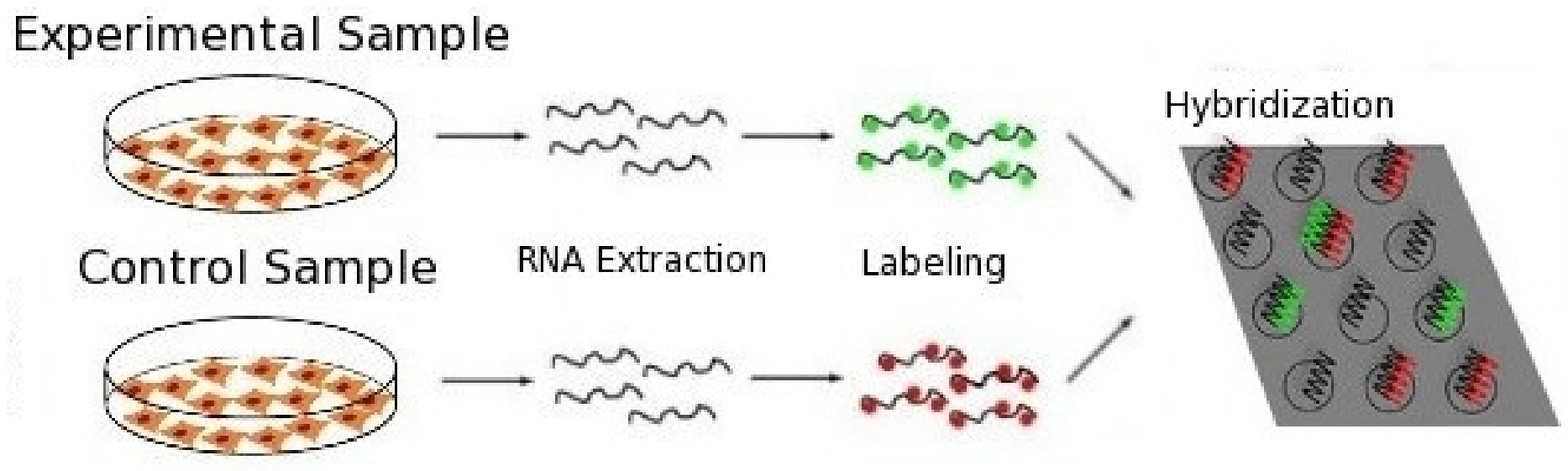}}
  \caption[Typical microarray experiment]{Illustration of a typical
    microarray experiment (using cDNA technology). First, mRNA is
    extracted from two groups of cells, namely an experimental sample
    of interest and a control sample. Each sample is labeled with a
    different color of fluorescent dye. The samples are then combined
    and hybridized onto an array. The relative abundance of the mRNA
    corresponding to a particular gene can be measured by calculating
    the ratio of red dye to green dye at the appropriate spot on the
    array.}
  \label{F:array_fig}
\end{figure}

\begin{figure}
  \centerline{\includegraphics[width=\textwidth]{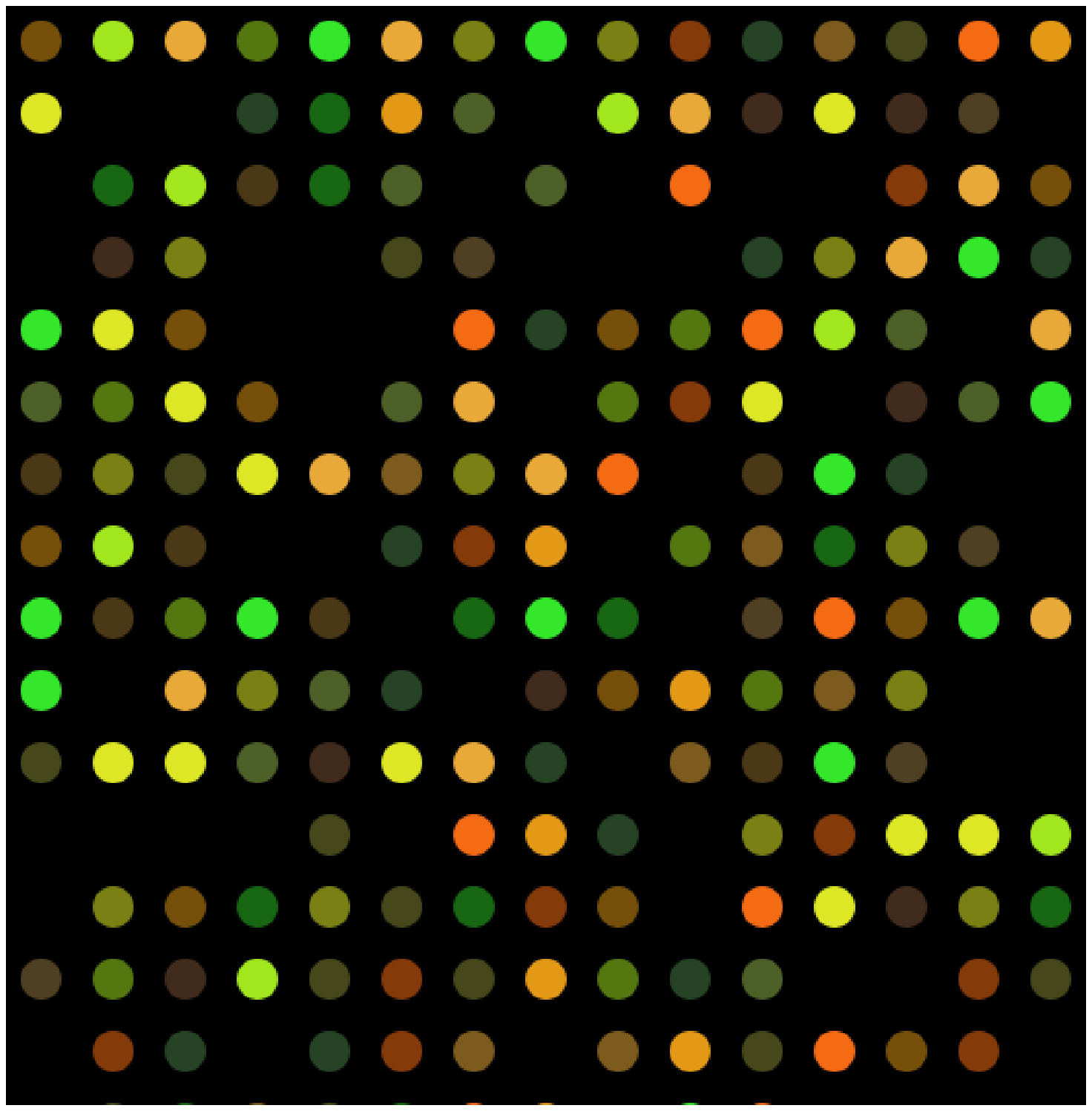}}
  \caption[DNA microarray slide]{Image of a DNA microarray slide. One
    may measure the relative gene expression of each gene by comparing
    the ratio of the amount of red dye to the amount of green dye at
    each probe on the array.}
  \label{F:microarray}
\end{figure}

Figure \ref{F:array_fig} illustrates a typical (cDNA) microarray
experiment. Two samples are collected, namely an experimental sample
and a control sample. For example, the experimental sample may contain
tissue from a cancerous tumor, and the control sample may contain
non-cancerous tissue from the same location in the body. First, mRNA
is extracted from both samples. The extracted mRNA is treated with an
enzyme called reverse transcriptase to convert it to a complementary
DNA (or cDNA) sequence. Then each sample is treated with a fluorescent
dye. Typically the red dye Cy5 and the green dye Cy3 are used. Equal
amounts of the two samples are then hybridized onto an array. To
determine which genes are expressed at a high (or a low) level in the
experimental group compared to the control group, one may measure
the ratio of Cy5 to Cy3 at the probe on the array corresponding to
that gene. For example, suppose that the experimental sample was
treated with the red dye and the control sample was treated with the
green dye. Then a red spot on the array indicates that there was more
mRNA for that particular gene produced by the experimental group than
the control group, indicating that this gene is expressed at a higher
level in the control group. An image of a DNA microarray slide is
shown in Figure \ref{F:microarray}.

\begin{figure}
  \centerline{\includegraphics[width=\textwidth]{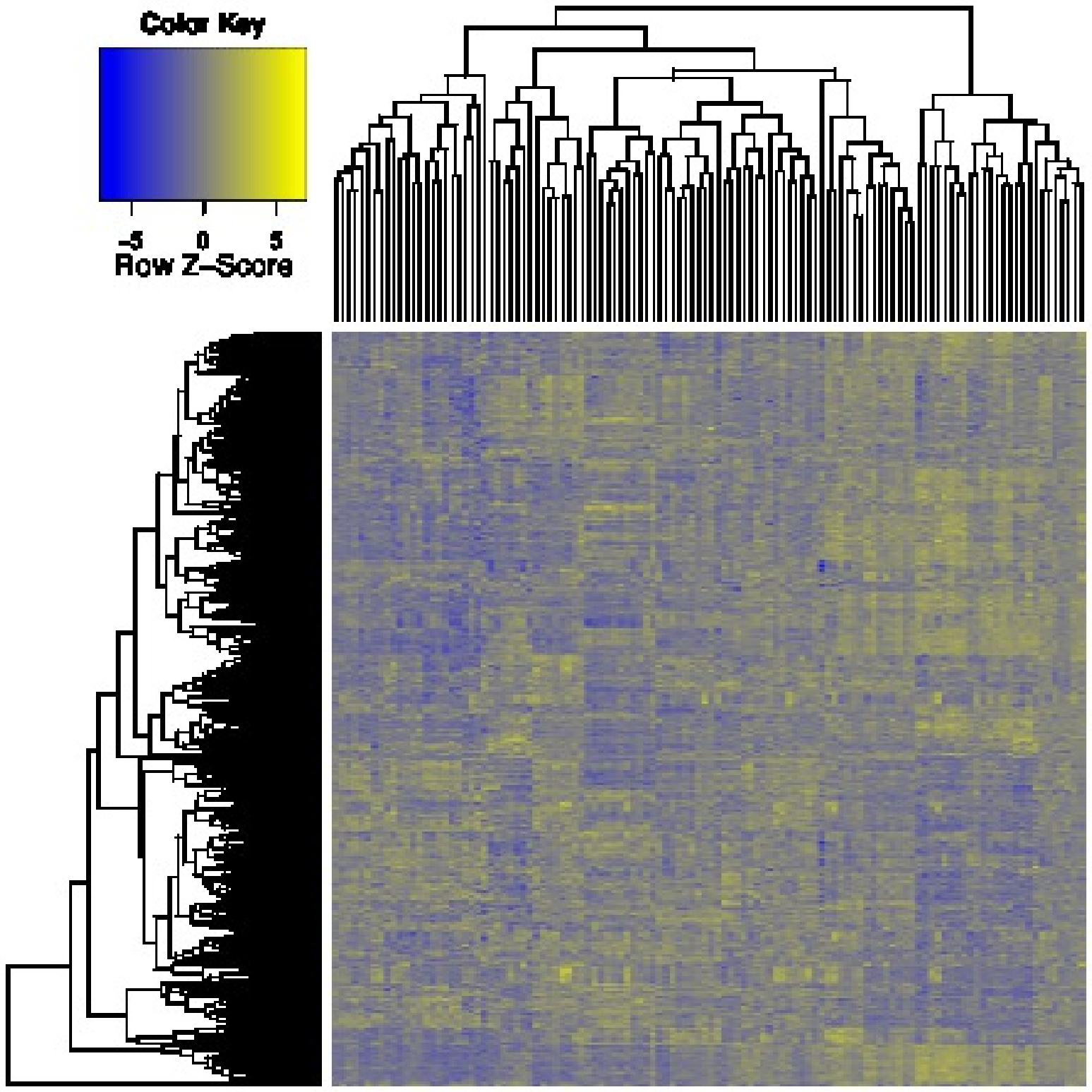}}
  \caption[Example of a microarray heat map]{Heat map of the
    leukemia microarray data of \citet{LB04}. Each colored square on
    the map corresponds to the expression level of a given gene for a
    given patient. In the above figure, each row represents a gene and
    each column represents a patient. The brighter the color of a
    given square, the higher (or lower) the expression level of the
    corresponding gene. Usually hierarchical clustering is performed
    on the rows and columns of the data set prior to drawing the heat
    map.}
  \label{F:heatmap}
\end{figure}

Before microarray data is analyzed, the ratio of red dye to green dye
at each spot on the array is measured using an appropriate
scanner. This ratio is stored in a large data matrix. Typically each
row of the data matrix contains all the measured expression levels for
a given gene, and each column of the data matrix corresponds to a
particular sample. Such a data set is often visualized in the form of
a heat map, as shown in Figure \ref{F:heatmap}. The task of the
microarray data analyst is to answer the biological question(s) of
interest using this data matrix.

It is important to attempt to remove extraneous variation in
microarray data prior to data analysis. Variations in design of the
arrays, sample preparation and scanner reading can produce ``batch
effects'' where some subset of samples exhibit systematic differences
in gene expression that are unrelated to the biological process of
interest. Failure to account for such batch effects can result in
spurious findings. \citep{BCN08, BC09} Thus, normalization is often
necessary to remove batch effects. Numerous methods have been proposed
to normalize microarray data \citep{gT01, yY02, jQ02, SS03, bB03,
rI05, jB08, pS08, jL10}. A detailed description of these normalization
methods are beyond the scope of this review; see the aforementioned
references for more information.

\section{Methods for Identifying Significant Features in DNA
  Microarray Data}
Perhaps the most common objective of microarray experiments is to
identify genes that are associated with a biological process of
interest. For example, one may wish to identify genes associated with
a disease of interest by comparing the expression levels of genes in
diseased samples to the corresponding expression levels in healthy
samples. Other outcomes of interest are also possible. For example, in
cancer studies, one frequently wishes to identify genes associated
with the survival time of cancer patients. The motivation is that
genes that are associated with lower survival are likely to be
associated with more serious forms of cancer that require more
aggressive treatment.

In statistical terms, one has a large number of features (genes) and
an outcome variable (e.g. disease versus control, survival time,
etc.). The objective is to identify genes that are associated with the
outcome variable. In principle, this objective can be accomplished
using conventional statistical methods. To compare the expression of
genes between two groups, one may calculate a $t$-test statistic for
each gene. If there are three or more groups, an ANOVA F-test
statistic may be used. To find genes associated with a continuous
outcome variable, one may calculate a standardized regression
coefficient, and to find genes associated with a survival outcome, one
may calculate the Cox score for each gene \citep{dB02, BT04}.

However, these conventional methods often perform poorly on microarray
data sets for several reasons, which will be discussed in more detail
below. DNA microarray data sets are frequently noisy, and sample sizes
are often small. Moreover, the gene expression levels are often highly
correlated with one another, and failing to account for this fact may
result in a loss of power. Also, one will typically perform several
thousand hypothesis tests in a microarray experiment, so specialized
methods are needed to control for type I error.

Throughout the remainder of this section, we will assume that one is
comparing two different conditions using a $t$-test or a variation of
the conventional $t$-test. However, the methods discussed below are
easily generalized to other test statistics, such as ANOVA F-tests,
standardized regression coefficients, and Cox scores.

\subsection{Fold Change Methods}
One simple method for identifying differentially expressed features is
to compute the average value of each feature under each condition and
then compute the ratio of these averages. If the ratio exceeds some
arbitrary cutoff, then the difference is called ``significant.'' For
example, a gene may be called ``significant'' if the average
expression level of a gene is more than twice as large (or less than
half as large) in one condition compared to the other.

This approach has the benefit of simplicity, and it has been used in
previous microarray studies \citep{mS96, jR97}. However, this method
has some serious shortcomings. It is not based on a formal statistical
test, so there is no simple way to calculate a $p$-value or confidence
interval or other measure of the statistical validity of the
association. Moreover, it is easy to see that this fold change has
higher variance for genes expressed at lower levels, which is true of
the majority of genes in microarray studies \citep{dR01, mN01,
vT01}. For these reasons fold change methods are generally accepted to
be inferior to other methods for identifying differentially expressed
features \citep{yC97, rM01, vB03, aH04, dA06}.

\subsection{$T$-Tests}
An alternative approach is to identify significant genes based on a
two-sample $t$-test of the null hypothesis that the mean expression
level of the gene is the same under both conditions. This approach has
also been used in microarray studies \citep{mC00}, and it has several
advantages over fold change methods. It is straightforward to
calculate $p$-values and confidence intervals using $t$-tests, and for
large samples the distribution of the $t$-statistic is independent of
the overall expression level of the gene. In contrast, fold change
statistics have higher variance for genes expressed at low levels.

Unfortunately, using $t$-tests to identify differentially expressed
genes can be problematic when the sample size is small, which is
commonly the case in microarray experiments. It can be difficult to
obtain accurate estimates of the variances of each group when the
sample sizes are small. In particular, if the the estimated variance
of a gene is small, which occurs frequently when a gene is expressed
at low levels \citep{vT01}, then the gene may have a large
$t$-statistic even if the fold change is small.

\subsection{Alternative Versions of $T$-Tests} \label{S:alt_ttest}
Given the shortcomings of $t$-tests described above, numerous authors
have proposed alternative versions of $t$-tests for identifying
significant features in gene expression data. Typically these methods
combine data from all the genes to obtain a regularized estimator of
the variance of a particular gene. In general, such variance estimates
are biased. However, since the usual estimator of the variance has
high variance when the sample size is low, a biased estimator of the
variance may have lower prediction error than an unbiased estimator,
since these biased estimators have lower variance than the unbiased
estimator. This is especially true when the sample size is small. See
\citet{HTF09} for a more detailed discussion of this phenomenon, which
is commonly referred to as the ``bias-variance trade-off.''

\begin{figure}
  \centerline{\includegraphics[width=\textwidth]{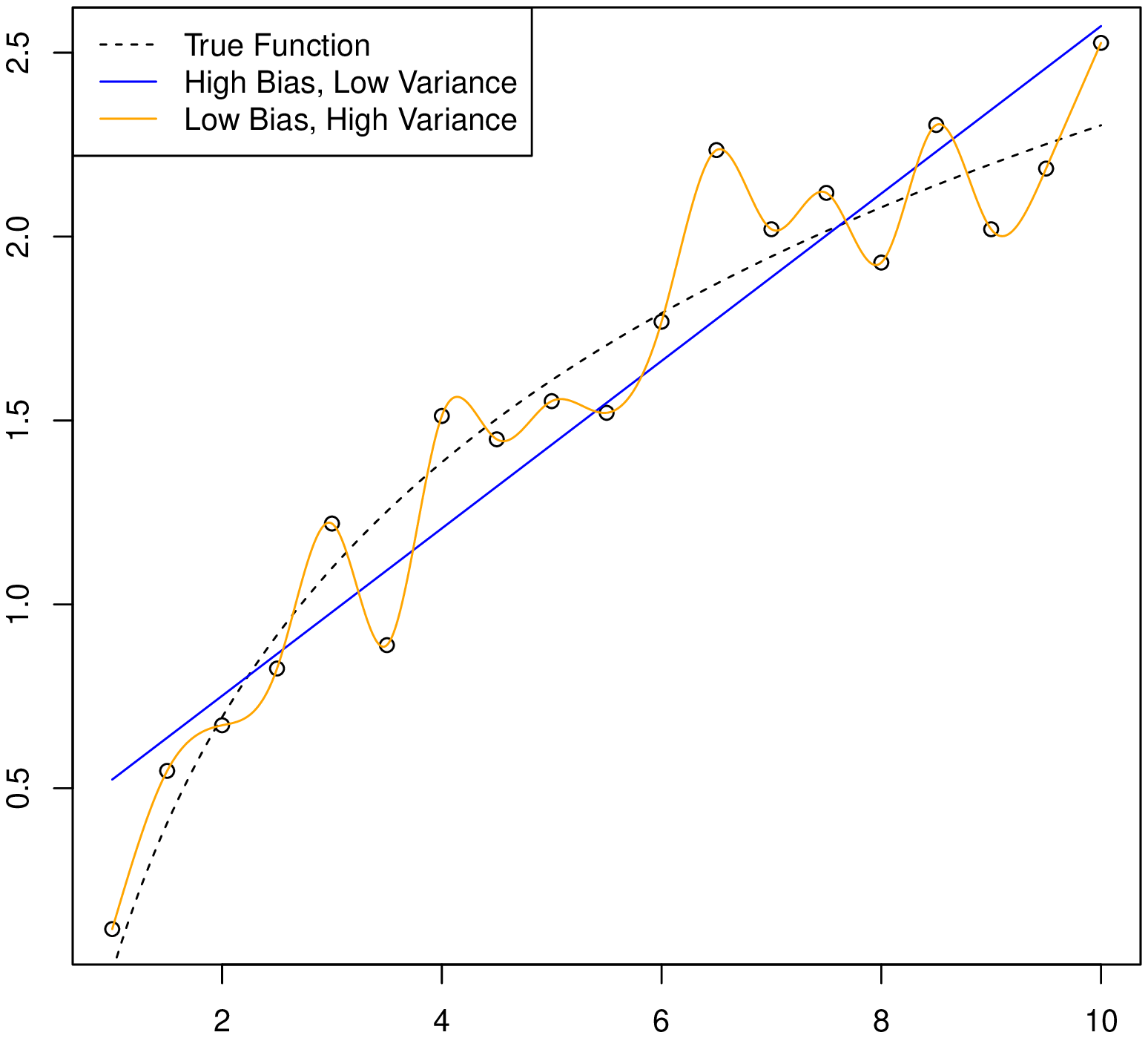}}
  \caption[Illustration of the bias-variance trade-off]{Illustration of
    the bias-variance trade-off. The above figure shows a regression
    problem where the objective is to predict $y$ given a value of
    $x$. The dotted line shows the true relationship between $x$ and
    $y$. The linear regression estimator (shown in blue) has high bias
    and low variance, and the interpolation estimator (shown in orange)
    has low bias and high variance.}
  \label{F:bv_spline}
\end{figure}

An example of the bias-variance trade-off is shown in Figure
\ref{F:bv_spline}. Suppose the objective is to predict $y$ based on
$x$ given the data in the figure. If one predicts $y$ using a linear
regression estimator based on $x$, the variance will be relatively low,
but the bias will be high, since it cannot model the nonlinear
relationship between $x$ and $y$. At the other extreme, if one
predicts $y$ by interpolating the data with a smoothing spline, the
bias will be 0, since the interpolation function can model any
arbitrary relationship between $x$ and $y$. However, the variance will
be high, since the predicted values of $y$ may change drastically if
such a model is fit to a new data set.

\begin{figure}
  \centerline{\includegraphics[width=\textwidth]{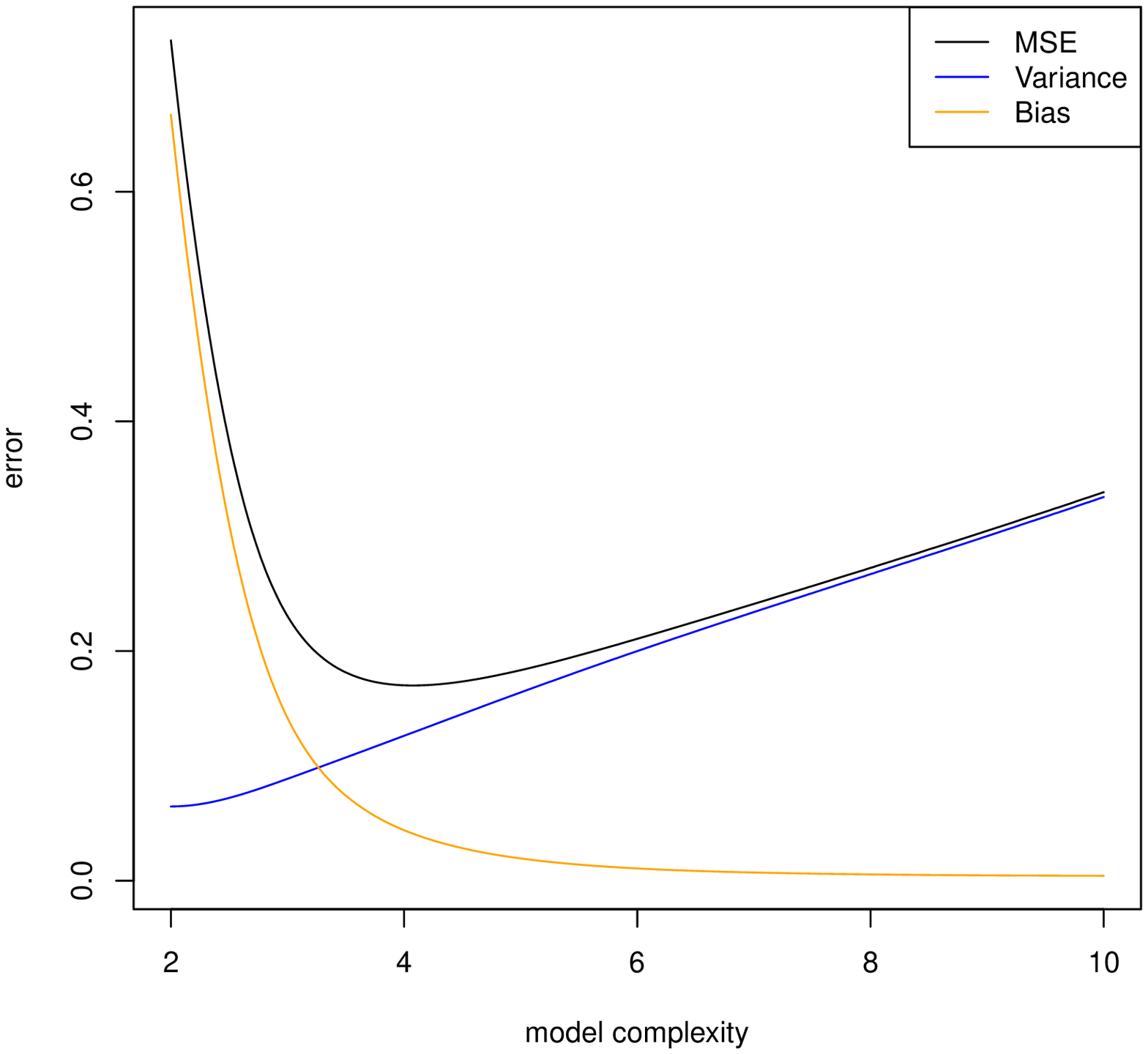}}
  \caption[Association between model complexity and
  bias/variance]{Illustrates the association between the complexity of
    a model and the bias/variance of the model. In general, as the
    complexity of a model increases, the variance of the model
    increases and the bias of the model decreases.}
  \label{F:bv_calc}
\end{figure}

Figure \ref{F:bv_calc} shows how the bias and variance of a series of
models varies as the complexity of the model increases. Each model in
this figure represents a smoothing spline \citep{HT90} fit to the data
from Figure \ref{F:bv_spline}. As the complexity of the model
increases, the variance of the model increases and the bias of the
model decreases. One attempts to choose the model complexity that
minimizes the expected prediction error or mean squared error (MSE),
which can be shown to be equal to the sum of the variance, the square
of the bias, and an irreducible error term due to unexplainable
variance in $y$ \citep{HTF09}.

These figures illustrate why a regularized (i.e. biased) estimator of
the variance of the genes may produce better results when identifying
significant features based on microarray data. By regularizing the
estimates of the variance, the complexity of each individual model is
reduced, increasing the bias of the model but decreasing the
variance. If the decrease in variance is sufficient to offset the
increase in bias, the accuracy of the overall model may be increased.

One possible approach is to estimate the variance of each gene by
using the pooled estimator of the variance of all genes. Although this
method has been used for several microarray studies \citep{tT00,
sA00, mK00}, it also has some serious shortcomings. This obviously
assumes that the variance of the expression levels of all genes are
approximately the same, which is unlikely to be true in most
situations. More importantly, since the denominator of the $t$-test will
be the same for all genes, this method is essentially equivalent to
the fold change method, since it selects the genes with the largest
mean differences without regard for the variance of an individual gene
and thus suffers from the same drawbacks as fold change methods. In
terms of the bias-variance trade-off, this pooled variance estimate has
low variance but high bias.

An alternative approach is to combine the variance estimator of each
gene with some sort of pooled estimator of the variance across the
genes. This avoids the high variance that results from estimating the
variance of each gene individually as well as the high bias that
results from relying entirely on a pooled variance estimate. For
example, the ``Significance Analysis of Microarrays'' (SAM) procedure
of \citet{vT01} uses the following test statistic:
\begin{equation} \label{E:sam_ttest}
  t_i = \frac{\bar{X}_i - \bar{Y}_i}{s_i+s_0}
\end{equation}
Here $t_i$ represents the $t$-statistic for the $i$th gene, and
$\bar{X}_i$ and $\bar{Y}_i$ represent the mean expression level of the
gene under each experimental condition. The variance is estimated by
summing the estimated variance of the $i$th gene (denoted by $s_i$)
and a normalizing constant $s_0$. This normalizing constant reduces
the variance of the estimator of the variance and hence reduces the
likelihood of obtaining false positive findings as a result of genes
whose estimated variance is small. Typically $s_0$ is chosen to be
some quantile (such as the median) of the $s_i$'s across all of the
genes. The SAM software is publicly available as an add-in for
Microsoft Excel (http://www-stat.stanford.edu/~tibs/SAM/). It is also
implemented in the ``samr'' R package.

Other normalized estimators of the variance of microarray samples have
been proposed. For example \citet{wH02} apply an
$\operatorname{arsinh}$ transformation to the gene expression data
that is designed to produce stable variance estimates irrespective of
the gene's overall expression level. This method is implemented in the
``vsn'' R package (available through the Bioconductor project at
http://www.bioconductor.org). \citet{xC05} combine gene-specific and
between-gene variance estimates using a James-Stein estimator
\citep{cS62}. R code for implementing this method is available at\\
http://www.stjuderesearch.org/depts/biostats/documents/cui-Fstat.R. In
general, any method to reduce the variance in the estimates of the
variances of the individual genes can produce more accurate results
when the sample size is small.

\subsection{Bayesian Methods}
Bayesian methods can also be used to combine information across genes
to avoid inaccurate variance estimates as a result of small sample
sizes. Typically these methods impose some type of Bayesian prior
distribution on the gene expression data and estimate the posterior
distribution for each gene by combining information across all of the
genes. For example, \citet{BL01} impose a prior distribution on the
variances of the genes to obtain the following regularized $t$-test:
\begin{equation} \label{E:bl_ttest}
  t_i = \frac{\bar{X}_i - \bar{Y}_i}{\sqrt{\frac{v_0 \sigma_0^2 +
        (n-1) s_i^2}{v_0+n-2}}}
\end{equation}
In this expression $\bar{X}_i$, $\bar{Y}_i$, and $s_i$ are defined as
they were in (\ref{E:sam_ttest}). The parameter $\sigma_0^2$ is an
estimator of the pooled variance across genes, which is calculated
using data from all the genes, and $v_0$ is a tuning parameter that
controls the relative contributions of the gene-specific variance
estimate and the global variance estimate. R code for implementing
this method is available at
http://molgen51.biol.rug.nl/cybert/help/index.html.

Note that (\ref{E:bl_ttest}) is similar to (\ref{E:sam_ttest}) in that
the denominator of the $t$-statistic consists of a linear combination
of an estimator of the variance of gene $i$ plus a pooled estimate of
the variance of all the genes. The similarity between the two
expressions is not surprising. In general Bayesian methods tend to
produce biased parameter estimates, but these estimators may have
lower variance/mean squared error than unbiased estimators, which is
the same motivation for considering the regularized variance
estimators discussed previously. Indeed, in some situations
regularized frequentist parameter estimators can be shown to be
Bayesian estimators with the appropriate choice of prior
\citep{mG76}.

Other similar Bayesian approaches have also been proposed for
different types of microarray problems \citep{LS02, cK03, WS03, gS04,
mN04}. In particular, the ``limma'' method of \citet{gS04} uses an
empirical Bayes test statistic that consistently performed well in a
recent study comparing feature selection methods for microarray data
\citep{JHC06}.

\subsection{Calculating $P$-Values}
If a $t$-test (or other conventional parametric test, such as ANOVA or
regression) is used to test the null hypothesis of no association
between the expression level of a given gene and an outcome, then
calculating the $p$-value for this null hypothesis is straightforward
if the assumptions of the test are satisfied. However, it may be
dangerous to assume that these test statistics are normally
distributed when the sample size is small. Moreover, as discussed
previously, in many situations it is preferable to use biased
estimators of the variance of a gene's expression level. When a biased
estimator of the variance is used, a $t$-statistic may no longer have
a $t$ distribution. Thus, alternative approaches may be needed to
compute $p$-values in these situations.

One possible alternative is to calculate $p$-values based on the
permutation distribution of the test statistic. Let $t_j$ denote the
$t$-statistic (or other test statistic) associated with gene
$j$. Suppose the sample labels are then permuted $K$ times, and let
$t_{j,k}$ denote the test statistic associated with gene $j$ for the
$k$th permuted data set. Then one can estimate the $p$-value for gene
$j$ (denoted by $p_j$) as follows:
\begin{equation} \label{E:perm_pvals}
  p_j = \frac{1}{K} \sum_{k=1}^K I\left(|t_{j,k}|>|t_j|\right)
\end{equation}
Here $I(x)$ denotes an indicator function that is equal to 1 if the
condition is true and 0 otherwise. In other words, the $p$-value is
estimated by counting the number of times that the permuted version of
the test statistic is ``more extreme'' than the original (unpermuted)
version of the test statistic. A very large (or very small) test
statistic is unlikely to occur by chance, so very few permuted data
sets will produce a larger test statistic and the $p$-value will be
small. This approach is used by the ``SAM'' software package
\citep{vT01} to calculate $p$-values.

This approach requires a choice of the number of permutations $K$. For
small data sets, one may simply evaluate all possible permutations. In
the case where one wishes to compare $n_1$ samples from one condition
to $n_2$ samples from another condition using a $t$-test (or variant
thereof), there are a total of $\binom{n_1+n_2}{n_1}$ possible
permutations. However, this would be computationally intractable for
larger data sets, so it is common to arbitrarily select a value of
$K=1000$ or an even larger number if more precision is desired.

One possible problem with calculating $p$-values using
(\ref{E:perm_pvals}) is that it can be difficult to estimate
$p$-values that are close to 0. If $|t_j|>|t_{j,k}|$ for all $k$, then
(\ref{E:perm_pvals}) implies that $p_j=0$, which in reality all that
can be inferred is that $p_j<1/K$. This is problematic because certain
methods for adjusting for multiple hypothesis testing in microarray
experiments require precise estimation of $p$-values that are very
close to 0. See below for more details.

There are a few possible solutions to this problem. The simplest
approach is to increase the value of $K$. This will solve the problem
given sufficient computing power, but it can be computationally
intractable for large data sets. Another possibility is to pool the
results of all the genes when calculating the permutation
$p$-values. Suppose there are a total of $N$ genes in the
experiment. Then we estimate $p_j$ as follows:
\begin{equation} \label{E:perm_pvals2}
  p_j = \frac{1}{NK} \sum_{i=1}^N \sum_{k=1}^K I\left(|t_{i,k}|>|t_j|
  \right)
\end{equation}
In other words, rather than simply counting the number of times that
the permuted test statistic for gene $j$ is more extreme than the
unpermuted test statistic for gene $j$, one counts the number of times
that the permuted test statistic for any gene is greater than the
permuted test statistic for gene $j$. This can increase the precision
of the estimates of $p_j$ without increasing the computational
burden. See \citet{HTF09} for a complete discussion of calculating
$p$-values based on the permutation distribution of the test
statistics.

\subsection{Methods for Combining Information Across Genes}
The methods discussed thus far assume that hypothesis tests will be
performed on each gene one at a time, and that the results of a
hypothesis test on a given gene will not be affected by the hypothesis
tests performed on other genes. This strategy may be inefficient on
DNA microarray studies. Genes often act in pathways, meaning that
several genes may be involved with the same biological process and
hence be activated and deactivated simultaneously. If several related
genes show evidence of differential expression at the same time, that
is stronger evidence that the differential expression represents
biological signal than if such a pattern were observed for a single
gene. Several methods have been proposed for combining information
across genes when searching for differentially expressed genes in
microarray studies, which will be discussed below.

\subsubsection{Biologically Motivated Methods}
One approach for combining information across genes is to utilize
known biological relationships among the genes. Typically genes are
classified into groups using biological databases such as Gene
Ontology \citep{mA00}. Each group represents a set of biologically
similar genes. The most commonly used methods compare the number of
significant features in each group to the number expected if the genes
in the group are not differentially expressed. If there are an
unusually high number of significant features in a given group, that
suggests that the pathway corresponding to the group is differentially
expressed.

One strategy for identifying pathways containing differentially
expressed genes is known as over-representation analysis (ORA). ORA
first identifies a list of ``significant'' genes using any of the
previously described methods for detecting differentially expressed
genes. The $M$ ``most significant'' genes are selected, which are
typically the genes with the smallest $p$-values. Then for each group
of genes, Fisher's exact test (or some approximation thereof) is used
to test the null hypothesis that the number of genes called
significant in each group does not exceed the number of genes expected
to be called significant due to chance. Various implementations of ORA
have been proposed in the literature \citep{kD02, sD03b, bZ03, sD03,
sZ03, gB03}, and it has been used in some microarray experiments
\citep{eB03}. 

Despite the popularity of ORA, it has several shortcomings. Typically
only the top $M$ genes are used to compute the ORA statistics,
resulting in the loss of any information available from genes not
among the $M$ most significant genes. The choice of $M$ is often
arbitrary as well. Moreover, all of the top $M$ genes are treated
equally, meaning that genes with extremely small univariate $p$-values
are given the same weight as genes whose univariate $p$-values are
much larger. Finally, ORA considers the gene to be the unit of
analysis rather than the subject, which is inappropriate in virtually
all real-world situations. Among other issues, it implies that the
gene sets should be independent of one another, which is almost
certainly not true in practice. See \citet{pP04}, \citet{lT05}, or
\citet{dA06} for more information on the shortcomings of ORA.

An alternative strategy that avoids the problems associated with ORA
is gene set enrichment analysis (GSEA). GSEA functions as follows:
First, the genes are ordered according to their $t$-statistics or
$p$-values or some other measure of univariate statistical
significance. Then for each group of genes, the distribution of the
$t$-statistics of the genes in the group is compared to the distribution
of the genes not in the group using a one-sided Kolmogorov-Smirnov
statistic or some other similar statistic. The idea is that if a
group of genes is differentially expressed, then the distribution of
the $t$-statistics among that set of genes should be different than the
distribution of the $t$-statistics among the remaining genes. A
$p$-value can be calculated for each set of genes by permuting the
data multiple times and using (\ref{E:perm_pvals}) or
(\ref{E:perm_pvals2}) or alternative methods. Various
implementations of GSEA have been proposed in the literature
\citep{vM03, rB04, jR04, pP04, wB05, lT05, aS05, jZ06, mN07,
ET07}. There are also several software implementations of GSEA. For
example, the Broad Institute offers software to perform GSEA in both
Java and R (http://www.broadinstitute.org/gsea/index.jsp) and a
variant of GSEA is implemented in the ``SAM'' software package.

The main shortcoming of GSEA is the fact that it tests a ``competitive
null hypothesis.'' Suppose we have two groups of genes, which we will
call gene group 1 and gene group 2. Then a smaller $p$-value for
testing the null hypothesis of no differential expression in gene
group 1 implies a larger $p$-value for testing this null hypothesis in
gene group 2 even if the expression levels in gene group 2 remain
unchanged. This occurs because the $p$-value for gene group 2 is
calculated by comparing the test statistics of the genes in gene group
2 to the test statistics of all genes not in gene group 2, including
the test statistics in gene group 1. Thus, if extreme test statistics
are observed in gene group 1, this decreases the significance of gene
group 2. See \citet{DG04} or \citet{dA06} for a more detailed
discussion of this phenomena. The development of methods for
identifying groups of genes associated with an outcome of interest
that avoids the shortcomings of ORA and GSEA is an active research
area.

\subsubsection{Statistically Based Methods}
Other strategies for combining information across genes use novel
statistical methods that do not require any knowledge of the
biological relationship between the genes. We have previously
discussed one possible statistical strategy for combining information
across genes, namely regularized or Bayesian estimators of the
variance of individual genes. By using information about the variance
of other genes to estimate the variance of a specific gene, the
variance of the test statistic is greatly decreased, and hence the
risk of false positives and false negatives is also decreased.
However, in recent years there several more advanced methods have been
proposed for combining information across genes which we will briefly
describe below.

\begin{figure}
  \centerline{\includegraphics[width=\textwidth]{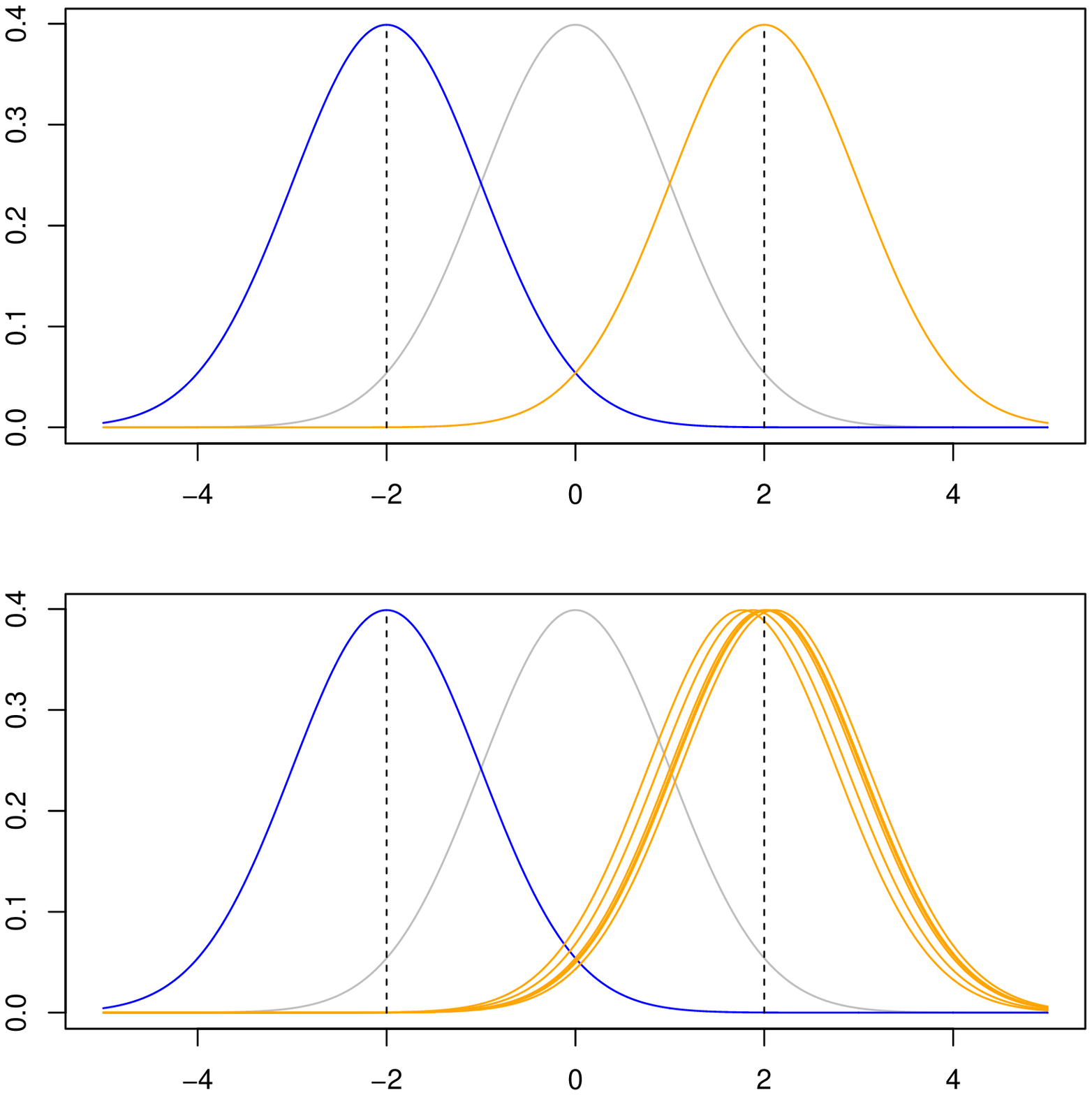}}
  \caption[Illustration of the ODP procedure]{Illustration of the ODP
    procedure. Suppose that the test statistic for the null hypothesis
    of no differential expression is $t=-2$ for one gene and $t=2$ for
    a second gene. Suppose further that there are several other genes
    with similar expression patterns to the second gene for which
    $t\approx 2$. Using traditional hypothesis testing procedures, one
    would be equally likely to reject the null hypothesis of no
    differential expression for both of the two genes. Using ODP, one
    would be more likely to reject the null hypothesis for the gene
    where $t=2$, since the existence of several genes with similar
    expression patterns increases ones confidence that the result is
    not due to chance.}
  \label{F:odp_fig}
\end{figure}

One strategy is known as the optimal discovery procedure (ODP)
\citep{jS07, SDL07}. The motivation for ODP is similar to the
motivation for the pathway-based methods discussed previously. Since
genes function in pathways, we expect that genes in the same pathway
are likely to be co-expressed. Thus, if a gene shows evidence of
differential expression, one can be more confident that the
differential expression is not due to chance if other genes show a
similar expression pattern. See Figure \ref{F:odp_fig} for an
illustration of this idea. The difference between ODP and
pathway-based methods is that pathway-based methods require one to
know in advance which genes are expected to be co-expressed based on
previously collected biological data whereas ODP does not.

The ODP is a generalization of the Neyman-Pearson lemma
\citep{NP33}. The Neyman-Pearson lemma states that the most powerful
test of a given null hypothesis against a given alternative hypothesis
rejects the null hypothesis when the ratio
\begin{equation} \label{E:np}
  \frac{\text{probability of the observed data under the alternative
      hypothesis}}{\text{probability of the observed data under the
      null hypothesis}}
\end{equation}
is large. The ODP generalizes the Neyman-Pearson lemma to situations
where multiple hypotheses are tested by rejecting the null hypothesis
that gene $i$ is not differentially expressed when the ratio
\begin{equation} \label{E:odp_np}
  \frac{\text{sum of the probabilities of observing data $i$ under
      each alternative hypothesis}}{\text{sum of the probabilities of
      observing data $i$ under each null hypothesis}}
\end{equation}
is large. Thus, if a set of genes with similar expression patterns all
show evidence of differential expression, then (\ref{E:odp_np}) will
be larger than (\ref{E:np}) for a given gene in the set, meaning that
the null hypothesis of no differential expression is more likely to be
rejected under ODP than under the traditional Neyman-Pearson
paradigm.

In practice, (\ref{E:odp_np}) cannot be computed exactly and must be
approximated. \citep{jS07, SDL07} Software for computing the ODP is
publicly available \citep{jL06}.

An alternative strategy for combining information across genes without
any biological information about the relationship between the genes is
the Lassoed Principal Components (LPC) method of \citet{WT08}. The
motivation for LPC is similar to the motivation for ODP: A gene is
more likely to be differentially expressed if there are other genes
with similar expression patterns than it is if there are no such
similar genes. However, LPC uses a different strategy to determine if
there are other genes with similar expression patterns. The idea
behind LPC is that if a group of genes are co-regulated, then it is
likely that a principal component of the gene expression matrix
(sometimes called an eigenarray \citep{APB00}) will capture the
variance in this group of genes. Thus, the LPC algorithm attempts to
identify an eigenarray or group of eigenarrays that are associated
with the biological process of interest and projects the $t$-statistics
(or other relevant test statistics) onto this group of
eigenarrays. This method can be shown to significantly reduce the
false discovery rate in a variety of situations. \citep{WT08} This
method is implemented in the ``lpc'' R package.

\subsection{Clustering and Prediction Methods}
Identifying features associated with an outcome of interest is not the
only objective of microarray studies. One may also wish to partition
the data into homogeneous subgroups and/or use the data to predict an
outcome of interest. Clustering methods and prediction methods are
useful in this situation.

There is a vast literature devoted to methods for clustering or
predicting an outcome based on microarray data. A full description of
such methods is beyond this scope of this review (which focuses on
feature selection). However, it is noteworthy that there are
methods for clustering \citep{XK01, jW03, BT04, TSV05, RD06, sK06,
wP06, PS07, BR08, mS08, WZ08, cM09, dK10, WT10} and prediction
\citep{rT02, BT04, nS04, eB06, bW06, TP07, WZ07, yG07, dP08, jG10,
SV11} that also perform feature selection. These methods generally do
not evaluate whether a selected gene is ``statistically significant''
nor do they indicate which genes are the ``most significant.'' Also,
the user of these methods often has limited control over the number of
features selected. Thus, these methods have serious disadvantages if
feature selection is the primary goal of the analysis. Nevertheless
these methods can identify a list of genes for further study,
particularly in cases where clustering and prediction are important
goals of the experiment.

\subsection{Comparison of Feature Selection Methods}
Numerous methods have been proposed for identifying significant
features in DNA microarray data. However, the question of which
methods produce the best results (i.e. maximize power while
controlling type I error) has not been studied extensively. In
practice researchers often choose feature selection methods based on
the ease of implementing the method rather than the performance of the
method. The ``SAM'' software package has become a popular tool for
microarray analysis largely due to the fact that it is available as an
Excel add-in and does not require the use of R or command-line
programs.

Limited research indicates that the ``limma'' method of \citet{gS04}
performs well for a wide variety of problems, although other methods
may perform better in specific situations  \citep{JHC06, WT08,
cM09b}. Limma is implemented in the ``limma'' R package, which is
available from Bioconductor. Determining which feature selection
method is likely to produce the best results on a given data set is an
important area for future research.

\section{Issues Related to Multiple Hypothesis Testing}
Identifying significant genes in microarray studies requires
performing a large number of hypothesis tests, which presents
statistical challenges. When performing a single hypothesis test, it
is conventional to choose a significance level $\alpha$ such that the
probability of rejecting the null hypothesis when it is true is equal
to $\alpha$. However, when multiple hypothesis tests are performed,
the probability of at least one false positive test will be much
larger than $\alpha$. Thus, methods are needed to control the number
of false positive tests while maintaining sufficient power to identify
truly significant genes.

\subsection{The Family-Wise Error Rate} \label{S:multtest}
One possible solution is to control the family-wise error rate (FWER)
at a specified level. The FWER is defined to be the probability of
rejecting at least one null hypothesis that is true. The most common
way to control the FWER at a specified level is to use a Bonferroni
correction: Each individual null hypothesis is rejected if and only $p
< \alpha/N$, where $p$ is the $p$-value for the test and $N$ is the
total number of tests. It is easy to show that the probability of at
least one type I error is no greater than $\alpha$ using this
procedure.

Although the Bonferroni correction controls the number of false
positive tests, it is a very stringent criterion that typically
results in a substantial loss of power. In experiments with small
sample sizes it is common for no tests to satisfy the Bonferroni
criteria. Thus, most microarray analysts prefer less stringent
approaches \citep{dA06}. Methods exist for controlling the FWER using
more permissive criteria than the Bonferroni correction \citep{sD02,
hW03}, but these methods also suffer from lower power and are not
commonly used.

\subsection{The False Discovery Rate}
The false discovery rate (FDR) is defined to be the expected
proportion of false positives among the set of genes that are called
significant. One may also adjust for multiple comparisons by
controlling the FDR rather than the FWER. This approach typically
yields greater power than FWER-based methods and hence is generally
regarded as preferable \citep{dA06}.

The FDR was first proposed by \citet{BH95}. To control the FDR at a
given level $\alpha$, they proposed the following procedure: Let
$p_{(1)} \leq p_{(2)} \leq \cdots \leq p_{(N)}$ be the ordered
$p$-values, and let $H_{(1)}, H_{(2)}, \ldots, H_{(N)}$ be the
corresponding null hypotheses. Then reject $H_{(1)}, H_{(2)}, \ldots
H_{(j)}$, where
\begin{equation} \label{E:bh_fdr}
j=\max_i \{i: p_{(i)} \leq \alpha i/N\}
\end{equation}
\citet{BH95} prove that the FDR of this procedure is at most
$\alpha$. This procedure is always valid if the $p$-values are
independent. It remains valid in some cases even when dependency
exists among the $p$-values, and methods exist for estimating the FDR
where any type of dependency exists. \citep{BY01, STS04, aF06, nM06,
PCP06, bE07, FDR07, LS08, RSW08, SC09, CH09, FKC09, FHG12}

Rather than choosing a specific FDR in advance, one may wish to
estimate the FDR when the top $m$ genes are called significant. This
is easy to do using the methodology of \citet{BH95}: If we let
\begin{equation} \label{E:fdr_est_bh}
  \hat{\alpha} = p_{(m)} N/m
\end{equation}
Then (\ref{E:bh_fdr}) implies that the FDR should be approximately
$\hat{\alpha}$.

If one estimates the null distribution of the test statistics using
permutations of the data as in (\ref{E:perm_pvals}) and
(\ref{E:perm_pvals2}), then an alternative estimator of the false
discovery rate may be used. Once again, let $t_j$ denote the
$t$-statistic (or other test statistic) associated with gene
$j$, and let $t_{j,k}$ denote the test statistic associated with gene
$j$ for the $k$th permuted data set. Also, let $t_{(1)} \leq t_{(2)}
\leq \cdots \leq t_{(N)}$ be the order statistics of the absolute
values of the $t_j$'s. Then one may estimate the FDR $\hat{\alpha}$
when the top $m$ genes are called significant as follows:
\begin{equation} \label{E:fdr_est_perm}
  \hat{\alpha} = \frac{1}{mK} \sum_{i=1}^N \sum_{k=1}^K
  I(|t_{i,k}| > t_{(N-m)})
\end{equation}
In other words, one estimates the FDR by dividing the average number
of genes called significant over $K$ permuted data sets by the number
of genes called significant in the unpermuted data set (which is $m$,
since the top $m$ genes were called significant). It can be shown that
$\hat{\alpha}$ in (\ref{E:fdr_est_perm}) is a consistent estimator of
the FDR \citep{jS02, STS04}. Also, it can be shown that estimators
(\ref{E:fdr_est_bh}) and (\ref{E:fdr_est_perm}) are equivalent
\citep{HTF09}. There are several R packages which will compute the FDR
using this methodology (such as the ``multtest'' package, which is
available from CRAN, and the ``fdrame'' package, which is available
from Bioconductor). This methodology is also implemented in the
``SAM'' software package.

\subsection{The Q-Value}
In multiple testing problems, the q-value \citep{jS03} of a given
test statistic $t$ is defined to be the smallest possible FDR that can
occur among all possible rejection regions that reject the null
hypothesis when $T=t$. For example, if a $t$-statistic is calculated
for each gene and the $j$th such $t$-statistic is $t_j$ and $|t_j|=C$,
then the q-value for the $j$th hypothesis test is the FDR for the
rejection region $|t_i| \geq C$. In other words, the q-value is the
FDR that results when one calls gene $j$ significant along with all
other genes that have a more extreme test statistic than gene
$j$. Obviously genes with more extreme test statistics will have
smaller q-values. The q-value may be estimated using
(\ref{E:fdr_est_bh}) or (\ref{E:fdr_est_perm}), although other
approaches are possible (see below). The q-value may be calculated
using the ``qvalue'' R package (available from Bioconductor) as well
as the ``SAM'' software package.

A Bayesian interpretation of the q-value is possible, as described in
\citet{jS02}, \citet{ET02} and \citet{jS03}. Suppose that each gene
comes from one of two populations, one of which consists of genes that
are differentially expressed, and the other which consists of genes
that are not differentially expressed. Under this assumption, the test
statistic for each gene may be modeled using a mixture model. Define a
set of random variables $Z_j$ such that $Z_j=0$ if gene $j$ is not
differentially expressed and $Z_j=1$ if gene $j$ is differentially
expressed. Also let $|t_j|=C$, and let $q(t_j)$ be the q-value
corresponding to gene $j$. Then one can show \citep{bE01, jS02} that
\begin{equation} \label{E:q_bayes}
  q(t_j) = P(Z_j=0 | |t_j| \geq C)
\end{equation}
In other words, under this mixture model, the q-value is the posterior
probability that the $j$th null hypothesis is true given the test
statistic for gene $j$. Although we have assumed a rejection region of
the form $|t_i| \geq C$ in (\ref{E:q_bayes}), this result holds under
more general rejection regions.

Note: (\ref{E:q_bayes}) is only true if we calculate the $p$-value
based on the positive false discovery rate (pFDR) as defined by
\citet{jS02} rather than the traditional FDR proposed by
\citet{BH95}. The pFDR is defined to be the expected proportion of
false positives among the set of genes that are called significant
conditional on the fact that at least one gene is called significant.

Methods exist for directly estimating the mixture distribution under
the model described above and thereby estimating the pFDR/q-value
corresponding to individual genes using (\ref{E:q_bayes}) or similar
procedures \citep{dA02, PM03, DMT05, bE08}. Limited research suggests
that many of these methods produce comparable results \citep{DD05,
dA06}. Such mixture models may also be used for omnibus testing.
\citep{DC08, DC10}

\begin{conclusion}
Microarrays have been important for a variety of biological
applications for over a decade. However, technology for generating
high-throughput biological data is improving at a rapid pace, and
technology that is commonly used today may be replaced in the near
future. Indeed, some have suggested that newer technologies such as
RNA-seq may soon replace microarrays \citep{jS08} just as microarrays
have largely replaced older techniques such as Northern blotting
\citep{mT01}. As technology advances, new methods will be necessary
for analyzing data sets generated by the new techniques, and some
methods for analyzing microarray data may no longer be useful in the
future if microarrays are replaced by newer methods. Indeed, methods
for identifying differentially expressed genes based on RNA-seq data
is currently an active research area. \citep{lW10, RO10, hK10, aH10,
jB10, LT11}

Despite these changing technologies, we feel that a discussion of
methods for analyzing microarray data is still relevant and
timely. DNA microarrays are still cheaper than RNA-seq assays, and
RNA-seq gene expression measurements can be unreliable for genes
expressed at lower levels \citep{pL11}. More importantly, however,
many of the statistical techniques that have been developed for
analyzing microarray data can also be applied to data produced by
other high-throughput biological assays. For example, using normalized
or Bayesian estimators of the variance of an estimator is useful for
performing feature selection in any situation where the number of
features is large and the number of observations is small. Similarly,
use of the FDR and pFDR to control type I error is useful for a wide
variety of multiple testing problems, which arise in the analysis of
nearly all types of modern high-throughput biological data. For
example, the ``SAM'' software package for DNA microarray analysis was
recently upgraded to analyze RNA-seq data in addition to DNA
microarray data. The new method continues to use resampling-based 
approaches to estimate the null distribution of each test statistic
which is then used to estimate the FDR. See \citet{LT11} for
details. Likewise, GSEA and other pathway-based methods for feature
selection have been applied to genome-wide association studies
\citep{iM09, kZ10, dN10}. Thus, we see that the methods developed for
DNA microarray analysis will be useful for many years in the future
even as technology changes.
\end{conclusion}

\begin{furread}
\citet{CC03} and \citet{dA06} provide good overviews of feature
selection methods for microarray data. \citet{JHC06} describe several
of the most commonly used feature selection methods for microarrays
and compare the performance of these methods on 9 publicly available
data sets. There are also numerous books containing information on
feature selection and other aspects of microarray data analysis not
considered in this review. Good references include \citet{CQB03},
\citet{gP03}, \citet{tS03}, \citet{WM04}, \citet{MDA05},
\citet{DMV06}, and \citet{sD11}.
\end{furread}

\section{Acknowledgments}
This work was partially supported by NIEHS grant P30ES010126 and NCATS
grant UL1RR025747. We thank the four anonymous reviewers for their
helpful suggestions.

\bibliographystyle{wires}
\bibliography{bibliography}

\begin{crossref}
\cref{CSA0004}, \cref{CSV0017}, \cref{CSA0034}
\end{crossref}

\endOverview
\end{document}